\documentclass[a4paper,11pt]{article}
\usepackage{amsmath}
\usepackage{graphicx}
\usepackage{amssymb}
\usepackage{amsfonts}
\usepackage{latexsym}

\def\beq{\begin{eqnarray}}
\def\eeq{\end{eqnarray}}
\def\ln{\,\mbox{ln}\,}

\def\Tr{\,\mbox{Tr}\,}

\def\al{\alpha}
\def\be{\beta}

\def\ga{\gamma}\def\de{\delta}

\def\ep{\epsilon}
\def\ze{\zeta}

\def\ka{\kappa}
\def\la{\lambda}

\def\pa{\partial}

\def\si{\sigma}
\def\om{\omega}
\def\ph{\varphi}

\def\Ga{\Gamma}

\def\La{\Lambda}

\newcommand{\eq}[1]{(\ref{#1})}

\newcommand{\nn}{\nonumber}
\usepackage{indentfirst}
\usepackage{mathrsfs} 

\usepackage{cite}
\usepackage{color}

\begin{document}

\renewcommand*{\thefootnote}{\fnsymbol{footnote}}

\begin{center}
{\large
On the Vilkovisky--DeWitt approach and renormalization group
\\
in effective quantum gravity}

\vskip 6mm
{\small \bf Breno L. Giacchini,}$^a$%
 \footnote{E-mail address: breno@sustech.edu.cn}
\quad {\small \bf Tib\'erio de Paula Netto}$^a$%
 \footnote{E-mail address: tiberio@sustech.edu.cn}
\quad and \quad {\small \bf Ilya L. Shapiro}$^{b}$%
\footnote{On leave from Tomsk State Pedagogical University.
\ E-mail address: shapiro@fisica.ufjf.br}
\vskip 6mm

(a) Department of Physics,
\ Southern University of Science and Technology,
\\ Shenzhen, \ 518055, \ China
\vskip 2mm

(b) Departamento de F\'{\i}sica, \ ICE, \
Universidade Federal de Juiz de Fora
\\ Juiz de Fora, \ 36036-900, \ MG, \ Brazil
\vskip 2mm

\end{center}
\vskip 4mm

\centerline{\uppercase{abstract}}
\vskip 2mm
\begin{quotation}

\noindent
The effective action in quantum general relativity is strongly
dependent on the gauge-fixing and parametrization of the quantum
metric. As a consequence, in the effective approach to quantum
gravity, there is no possibility to introduce the
renormalization-group framework in a consistent way.
On the other hand, the version of effective action
proposed by Vilkovisky and DeWitt does not depend on
the gauge-fixing and parametrization off-shell,
opening the way to explore the running of the cosmological and
Newton constants as well as the coefficients
of the higher-derivative terms of the total action.
We argue that in the effective framework the
one-loop beta functions for the zero-, two- and four-derivative terms
can be regarded as exact, that means, free from corrections
coming from the higher loops. In this perspective, the running
describes the renormalization group flow between the present-day
Hubble scale in the IR and the Planck scale in the UV.
\vskip 4mm

\end{quotation}

\renewcommand*{\thefootnote}{\arabic{footnote}}
\setcounter{footnote}{0}

\section{Introduction}
\label{Sec0}

The effective action in quantum field theory can be used for
deriving the $S$-matrix or other physically relevant quantities. In the
conventional approach, the effective action has fundamental
ambiguities related to the choice of parametrization of quantum
field or, in gauge theories, to the choice of the gauge-fixing
condition. These ambiguities vanish on-shell, which enables one to
formulate in a consistent manner the renormalizable theory.
However, this situation creates serious difficulties in the effective
theory. For instance, the renormalization group framework has to be
constructed based on the off-shell effective action. In many
cases this means that there is no way to use the renormalization group
to explore the running of the relevant parameters with scale in some
physical situations.

The quantum version of general relativity is non-renormalizable,
but it is perfectly appropriate for the effective theory approach.
The reason is that the graviton is a massless field, and the next
physical degrees of freedom (coming from higher derivatives) have
masses of the Planck order of magnitude $M_P \approx  10^{19}$ GeV
\cite{don}. However, in the case of quantum general relativity, the
one-loop and higher-loop divergences are strongly dependent on the
gauge-fixing and parametrization of the quantum metric. Owed to this,
the beta functions for all relevant parameters are badly defined and
there is no chance to extract the unambiguous running from the
effective low-energy quantum gravity.

The on-shell conditions in the non-renormalizable theory can hardly
be implemented, especially at the non-perturbative level.
Thus, it is difficult to obtain the relevant information from the quantum
theory.
On the other hand, there is
an alternative definition of the effective action, introduced by
Vilkovisky \cite{Vil-unicEA} and DeWitt \cite{DeWitt-ea},
based on the covariant formulation in the space of the quantum
gauge fields. The Vilkovisky--DeWitt unique effective action
is gauge- and parametrization-independent,
paving the way to explore in a consistent manner the
running of the parameters of the total action, including the
cosmological and Newton constants and the coefficients of the
higher-derivative terms.

The gauge-fixing independence of the Vilkovisky--DeWitt effective
action has been proved in a general setting and was also confirmed by
one-loop direct calculations~\cite{Bavi83-85,Fradkin:1983nw,Rebhan:1986wp,
Rebhan:1987cd,Ellicott:1987ir,
Kunstatter:1986qa,Huggins:1987zw}. Recently its universality
has been also confirmed by an explicit calculation in an arbitrary
parametrization within quantum gravity~\cite{UEA-BTS}.
Let us stress that this verification is especially relevant in the
case of conformal parametrization of the metric. One of the reasons
is that there is an exception in the general proof of universality,
which is mentioned in the pioneer work~\cite{Vil-unicEA}. It is
known that the Vilkovisky--DeWitt effective action depends on the choice of
the configuration space metric \cite{Huggins:1987zw}.
However, in the two-dimensional ($2D$) quantum gravity this metric
may depend on the choice of the gauge-fixing and therefore the
unique effective action becomes gauge-fixing dependent
\cite{Banin-PLB}. It was shown in  \cite{UEA-BTS} that this
effect does not take place in the four-dimensional ($4D$)
quantum gravity, regardless of the algebraic similarity with the
$2D$ case in the conformal parametrization. The important
detail is that the configuration-space metric has to be chosen
as the bilinear form of the classical action in the {\it minimal}
gauge, in a given parametrization of the quantum metric. In what
follows we shall base all considerations on this assumption,
which provides the universality of the Vilkovisky--DeWitt effective
action.

In the present work, we shall follow the previous publication on the
subject \cite{TV90} and consider the renormalization group equations
in the effective quantum gravity based on general relativity in the
unique effective action formalism. Indeed, we expand the analysis
performed in this seminal work in several directions. First of all,
we stress the importance of the effective approach and discuss in
more detail the corresponding area of application for the running of
the parameters. Second, we extend the analysis to the higher-derivative
part of the action, which is renormalized in a way similar to that
of the semiclassical gravity. Third, treating the
renormalization group in the effective framework, we argue that
the low-energy
one-loop running of the Newton constant $G$, cosmological constant
$\La$ and the parameters of the higher-derivative part of the action
is, in fact, non-perturbative and the corresponding beta functions
can be viewed as exact\footnote{We greatly appreciate the
contribution of the referee of the first version of the paper
\cite{UEA-BTS}, who gave us a valuable hint in this direction
and advised to proceed this discussion independently of the
parametrization-related calculations.}.

The manuscript is organised as follows. In the next Sec.~\ref{Sec2}
we briefly describe the general framework of the unique effective
action in quantum general relativity. The reader can consult the
parallel paper \cite{UEA-BTS} for further details. In Sec.~\ref{Sec3}
we construct, solve, and discuss the renormalization group equations
for the Newton and cosmological constants in the framework of
effective quantum gravity. In Sec.~\ref{Sec4} we extend  the analysis
to the higher-derivative sector of the theory and the discuss the
non-perturbative aspects of the corresponding running.
The perspectives of physical applications of the exact effective running
are briefly described in Sec.~\ref{Sec5}.
Finally, in Sec.~\ref{Sec6}, we draw our conclusions.

\section{Vilkovisky--DeWitt effective action in quantum gravity}
\label{Sec2}

Let us start by formulating general definitions, valid for any gauge
field theory, which we subsequently particularise for quantum gravity.
As mentioned before, off the mass shell the effective action depends
on the field parametrization. This can be readily seen by recalling that
in the one-loop approximation the effective action depends on Hessian
of the classical action. In fact, while the action
$S(\ph)$ is a scalar in the space of fields $\ph^i$, its second functional
derivative does not transform as a tensor. The dependence on the
gauge-fixing
can be understood in a similar manner, since the effective actions
calculated in different gauges are related by changes of
variables in the form of canonical transformations~\cite{Voronov:1982ur,Voronov:1984kq,Voronov:1982ph}.

The problem can be addressed in a geometric framework via the
introduction of a metric $\bar{G}_{ij}$ and a connection
$\mathscr{T}^k_{ij}$ in the configuration space of physical fields.
Accordingly, the definition of the effective
action should be modified, so that it is constructed
only with scalar quantities. This programme was carried out for the
first time\footnote{See Ref.~\cite{Honerkamp} for an earlier attempt
towards this in the context of non-linear $\si$-models.} by
Vilkovisky~\cite{Vil-unicEA}, who introduced the unique
effective action $\Ga(\ph)$ through
\beq
\exp i \Ga(\ph)
\,=\,
\int \mathcal{D} \ph' \mu(\ph^{\prime})
\,\exp\left\lbrace
i \left[ S(\ph^{\prime}) + \si^i(\ph,\ph^{\prime}) \Ga_{,i}(\ph) \right]
\right\rbrace,
\label{UEA_def}
\eeq
where  $\si_i(\ph,\ph^\prime)$ is the derivative (with respect to
$\ph^i$) of the world function \cite{BDW-65,J.L.Synge:1960zz}
constructed with the connection~$\mathscr{T}^k_{ij}\,$, and
$\mu(\ph^{\prime})$ is an invariant functional measure.
Since $\si^i(\ph,\ph^\prime)$ behaves as a vector with respect to
$\ph^{i}$ and as a scalar with regard to $\ph^{\prime i}$,
the effective action $\Ga (\ph)$ in~\eqref{UEA_def} is
reparametrization invariant and gauge independent.

The geometric objects mentioned above can be constructed from
two fundamental quantities: the
metric $G_{ij}$ in the space $\mathscr{M}$ of fields and the
generators $R^i_\al$ of gauge transformations.
As our main concern here is on gravity theory, we assume that
the field $\ph^i$ is bosonic, and that the generators $R^i_\al$
are linearly independent and form a closed algebra, with structure
functions which do not depend on the fields. Introducing the metric
$N_{\al\be}$ on the gauge group $\mathscr{G}$ and its inverse
$N^{\al\be}$,
\beq
N_{\al\be} \,=\, R^i_\al G_{ij} R^j_\be ,
\qquad
N_{\al\la} N^{\la\be} \,=\,\de^{\,\be}_\al,
\label{N_def}
\eeq
one can define the projector on $\mathscr{M}/\mathscr{G}$
\cite{Vil-unicEA,Fradkin:1983nw},
\beq
\label{Projetor}
P^i_j \,=\, \de^i_j - R^i_\al N^{\al\be} R^{k}_\be G_{kj}.
\eeq
The projected metric in the space $\mathscr{M}/\mathscr{G}$
of physical fields is, therefore,
\beq
\bar{G}_{ij}
\,=\,
P^k_i  G_{k\ell} P^\ell_j
\, = \, G_{ij} - G_{ik} R^k_\al N^{\al\be} R^\ell_\be G_{\ell j}.
\eeq
Since $N^{\al\be}$ is the inverse of a differential operator, this metric
contains a non-local part which arises from the constraints
imposed by the gauge symmetry.

The connection $\mathscr{T}^k_{ij}$ can be obtained by requiring its
compatibility with
$\bar{G}_{ij}$ (see \textit{e.g.}
\cite{Kunstatter:1986qa,Huggins:1987zw}),
and it can be written in the form~\cite{Vil-unicEA}
\beq
\mathscr{T}^k_{ij}
\,=\, \Ga^k_{ij} +  T^k_{ij},
\eeq
where $\Ga^k_{ij}$ are the (local) Christoffel symbols associated
to the metric $G_{ij}$ and
\beq
\label{Tijk}
T^k_{ij} \,=\, - 2 G_{(i|\ell} R^\ell_\al N^{\al\be}
\mathscr{D}_{|j)} R^k_\be
+ G_{(i|\ell} R^\ell_\al N^{\al\be} R^m_\be (\mathscr{D}_m R^k_\ga)
N^{\ga\de} R^n_\de G_{n|j)}
\eeq
is its non-local part. The parenthesis in the indices
denote symmetrization in the pair $(i,j)$ and $\mathscr{D}_i$
indicates the covariant derivative based on the Christoffel
connection $\Ga^k_{ij}$.

One can proceed the loop expansion of the Vilkovisky effective action~\eq{UEA_def},
\beq
\Ga (\ph) = S( \ph) + \bar{\Ga}^{(1)} (\ph)
+ \bar{\Ga}^{(2)} (\ph) + \cdots ,
\qquad\qquad
\hbar = 1,
\eeq
which gives the one-loop contribution~\cite{Vil-unicEA}
\beq
\label{EA1loop0}
\bar{\Ga}^{(1)}
\,=\,
\frac{i}{2} \Tr \ln G^{ik} \big(\mathscr{D}_k \mathscr{D}_j S
- T^\ell_{kj} S_{,\ell} - \chi^\al_{,k} Y_{\al\be} \chi^{\be}_{,j} \big)
- i \Tr \ln M^\al_\be,
\eeq
where
$\chi^\al$ is a gauge-fixing condition,
$Y_{\al\be}$ is the
weight functional and $M^\al_\be = \chi^\al_{,i} R^i_\be$ is the
Faddeev--Popov ghost matrix.
Compared to the standard effective
action, the Hessian of the classical action has now been replaced by
its covariant version, ensuring its tensor nature concerning field
reparametrizations. Notice also that the non-local part of the
connection \eq{Tijk} behaves as a tensor as well.
The divergent part of the one-loop effective action \eq{EA1loop0}
can be evaluated, \textit{e.g.}, by applying the generalized Schwinger--DeWitt
technique of Ref.~\cite{Bavi83-85}.

Nevertheless, the effective action defined by~\eqref{UEA_def} cannot be
viewed as a final solution aiming to the off-shell universality because of
two reasons. First, it might happen that the metric $G_{ij}$ is not uniquely
defined. This issue can be solved by additional prescriptions~\cite{Vil-unicEA}, as we shall
comment later. The most serious obstacle, however, is the lack of one-particle
irreducibility of the diagrams generated by~\eqref{UEA_def}, which may take
place beyond one-loop. Indeed, it gives rise to non-local divergences at the
two-loop approximation in Yang--Mills theories, as shown in Ref.~\cite{Rebhan:1986wp}.

The modification of the Vilkovisky effective action~\eqref{UEA_def} proposed by
DeWitt~\cite{DeWitt-ea} can be viewed as a way out of this difficulty, inasmuch
as it restores the one-particle irreducibility of the perturbative
expansion~\cite{Rebhan:1986wp,Rebhan:1987cd}
(see also~\cite{Ellicott:1987ir} for further discussion). For example,
in~\cite{Rebhan:1986wp} it was verified by explicit calculations that the
aforementioned non-local divergences which appeared in original
formulation~\eqref{UEA_def} in the two-loop Yang--Mills theory are cancelled
in the DeWitt approach~\cite{DeWitt-ea}.

The construction introduced by DeWitt~\cite{DeWitt-ea}, usually called
Vilkovisky--DeWitt effective action, consists in choosing an arbitrary
point $\ph^i_\ast$ in the configuration space and, instead of simply defining the
vector-scalar quantity $\si^i(\ph,\ph^\prime)$ in terms of the mean
field, one builds a system of Gaussian normal coordinates with $\ph^i_\ast$,
according to which the covariant Taylor expansions should be performed.
The method involves defining an effective action and a mean field
which depend on $\ph^i_\ast$, and only in the end, after the implicit
equation is solved iteratively, this point is identified to the mean field.
At one-loop level, the Vilkovisky--DeWitt effective action coincides to the
Vilkovisky one~\eqref{UEA_def}, therefore the Eq.~\eqref{EA1loop0} also
holds in this more general formalism.
Since for most of the discussion in the present paper we work with one-loop
results, we shall not present
further technical details of the Vilkovisky--DeWitt effective action. The
important point here is to stress that the formalism guarantees the
one-particle irreducibility of the diagrammatic expansion.

The remaining question to deal with is the aforementioned dependence of
the unique effective action and, in particular, its one-loop part~\eqref{EA1loop0},
on the choice of the metric in the space of the fields. This issue
is especially relevant for quantum gravity, where there is a one-parameter
family of such metrics, characterised by the parameter $\bar{a} \neq - 1/4$, given by
\cite{DeWitt:1967yk}
\beq
\label{MetricG}
G^{\mu\nu,\al\be}
\,=\,
\frac{1}{2} \, ( \de^{\mu\nu,\al\be}
\, + \, \bar{a} \, g^{\mu\nu} g^{\al\be} ),
\quad \quad \text{where} \quad \quad
\de^{\mu\nu,\al\be} = \frac{1}{2} ( g^{\mu\al} g^{\nu\be} + g^{\mu\be} g^{\nu\al} ).
\eeq
It was shown by explicit calculations \cite{Huggins:1987zw}
(see also \cite{Odintsov:1991yx,BarberoG.:1993cw}) that the
Vilkovisky--DeWitt effective action depends on the choice of $\,\bar{a}$.
Nonetheless, already in Ref.~\cite{Vil-unicEA} it was
introduced a prescription to fix this ambiguity. In accordance, the
field-space metric should coincide with the expression in the
highest-derivative term in the bilinear part of the classical action
in the minimal gauge fixing. In the parallel paper \cite{UEA-BTS}
we have shown that this prescription works perfectly well even
under changes of the parametrization of the quantum field, which
also modifies the parameter $\bar{a}$. Thus, it defines a unique
off-shell effective action.

The classical action of general relativity has the form
\beq
\label{action}
S\,=\, - \frac{1}{\kappa^2} \int \text{d}^4 x \sqrt{-g}(R + 2 \La),
\eeq
where $\,\ka^2 = 16 \pi G$ and $G$ is the Newton constant.
In the effective approach, the quantum theory takes into account
only the massless modes of the quantum metric \cite{don,Burgess}.
For this reason, these quantum effects are completely defined by
the action (\ref{action}), regardless of the presence of
higher-derivative terms (to be defined below) in the full gravitational action.

The one-loop divergent part of the Vilkovisky--DeWitt effective
action for the Einstein gravity was evaluated for the first time in
Ref.~\cite{Bavi83-85}, while the terms related to the cosmological
constant were calculated in~\cite{Fradkin:1983nw}. We shall skip
all the calculations and refer the reader to the mentioned papers for
the details. The result for the one-loop divergences is
\beq
\bar{\Ga}^{(1)}_{\text{div}}
\,=\,  - \frac{\mu^{n-4}}{\ep}
\int  \text{d}^n x \sqrt{-g} \,
\left\{ \frac{121}{60} C^2
- \frac{151}{180} E + \frac{31}{36} R^2
+ 8 \La R + 12 \La^2\right\}.
\label{Final-natural}
\eeq
where $\ep = (4\pi)^2(n-4)$  is the parameter of dimensional
regularization and $\mu$ is the renormalization parameter.
Also, $C^2
=R_{\mu\nu\al\be}^2 - 2R_{\mu\nu}^2 + \frac{1}{3}R^2$ denotes the
square of the Weyl tensor and
$E=R_{\mu\nu\al\be}^2 - 4R_{\mu\nu}^2 + R^2$ is the integrand
of the Gauss--Bonnet invariant in the four-dimensional spacetime.
On the classical mass shell, Eq.~\eqref{Final-natural} reduces to the
divergent on-shell part of the usual effective
action~\cite{hove,chrisduff}
\beq
\bar{\Ga}^{(1)}_\text{div} \big|_{\text{on-shell}}
\,=\, - \frac{\mu^{n-4}}{\ep} \int \text{d}^n x \sqrt{-g}
\left\{
\frac{53}{45} E -\frac{58}{5}  \La^2 \right\}.
\label{Final-onshell}
\eeq
This expression is also gauge-fixing and parametrization
independent (see \textit{e.g.} \cite{JDG-QG} and references therein),
but the advantage of the result \eqref{Final-natural} is that it is
universal even off-shell. This feature opens the way for formulating
consistent low-energy renormalization group equations for
$\La$, $G$ \cite{TV90}, and for other parameters of the action,
which were not included in the basic formula (\ref{action}).

Before  proceeding to the renormalization group and the effective
approach to quantum gravity, let us briefly review the power counting
in the quantum theory based on general relativity. In this theory, the
propagator behaves like $k^{-2}$ and there are two kinds of
vertices: the ones with two derivatives, owed to the
Einstein--Hilbert term, and those with no derivative, coming with
coefficient $\La$. The coupling constant (parameter of the loop
expansion) is $\ka^2$, with dimension of inverse of mass-squared. Since
the quantum metric $h_{\mu\nu}= g_{\mu\nu}-\eta_{\mu\nu}$ is
dimensionless, the power counting is especially simple. For a given
$p$-loop diagram with $n_2$ vertices with two momenta and
$n_0$ vertices with zero momenta, the superficial degree of
divergence is
\beq
\om \,=\,  2p -  2  n_0 + 2 - d,
\label{omega}
\eeq
where $d$ is the number of derivatives acting on the external
lines of $h_{\mu\nu}$. This relation defines the number of
derivatives $d=  2p -  2  n_0 + 2$ for the logarithmically divergent
diagrams with $\om=0$. It is easy to see that the expression
\eqref{Final-natural} satisfies this condition; the
$\mathcal{O}(\La)$-terms with $n_0=1$ are proportional to $R$,
that means $d=2$, while  $\mathcal{O}(\La^2)$-terms with $n_0=2$
have $d=0$.

Both at the one-loop level and
in higher loops, the  logarithmically divergent diagrams with $n_0=0$
satisfy the condition $d=  2p + 2$, which means that the maximal number
of derivatives in the counterterms grows linearly with the number
of loops $p$. In particular, for $\La=0$ the four-derivative
terms in the one-loop formula \eqref{Final-natural} are actually exact,
since they do not gain higher-loop contributions. The same concerns
the ${\cal O}(R^3_{...})$-type terms at the two-loop order, and so on.

On the other hand, in the real world, these terms are practically
exact even if $\La \neq 0$. The reason is that the four-derivative
terms gain $p$-loop contributions with coefficients proportional
to $\La \ka^2 = \frac{\La}{M_P^2}$ to the power $p$. In the
present-day Universe, this coefficient is of the order of $10^{-120}$,
which is small enough to support the argument that the result can be
regarded as exact.
It is a direct exercise to extend this statement also to the
lower-derivative terms in  \eqref{Final-natural}. We shall come back to
this reasoning and use it intensively in the next two sections when
discussing the renormalization group.

\section{Renormalization group based on the unique effective action}
\label{Sec3}

One can use the result~\eqref{Final-natural} for analyzing the renormalization group
equations in the low-energy (infrared, IR) sectors of the theory.
Such a construction has a direct physical sense. In the high-energy
domain (UV) the theory~\eqref{action} cannot be applied without
restrictions, as it is non-renormalizable
 and the contributions of
massive degrees of freedom, related to higher derivative terms,
are supposed to modify the beta functions.
However, since the quantum gravity based on general relativity is
a massless theory, it makes sense to explore the renormalization
group running in the IR. Assuming that the higher-derivative
 massive degrees of freedom
have masses of the Planck
order of magnitude, in  most of the physically relevant situations
these modes decouple \cite{frts82} (see also the concrete discussion
of this issue in the semiclassical gravity
\cite{apco,Franchino-Vinas:2018gzr} and qualitative
discussion in quantum gravity \cite{Gauss,SRQG-betas,324}),
such that the
running is completely defined by the action~\eqref{action}.

In other words, since the theory is massless, the quantum gravity
based on general relativity can be regarded as an effective theory
of quantum gravity at the energies between the UV (Planck) scale,
where the massive degrees
of freedom coming from higher derivatives can become relevant, and the
deep IR scale. Thus, the Vilkovisky--DeWitt unique effective action
enables one to explore the scale dependence in this vast region in a
gauge-fixing and parametrization independent manner.

From the classical action~\eqref{action} and the expression for
the divergences~\eqref{Final-natural}, it is easy to obtain the renormalization
relations
\beq
\frac{1}{\ka^2_0}\,=\,\mu^{n-4}\Big[
\frac{1}{\ka^2} - \frac{8}{(4\pi)^2(n-4)}\,\La\Big],
\qquad
\La_0\,=\,\La\Big[
1 + \frac{2}{(4\pi)^2(n-4)}\,\La \ka^2\Big].
\label{renrels}
\eeq
The bare quantities $\ka^2_0$ and $\La_0$ are $\mu$-independent,
as it is the case for the renormalized effective action. Applying the
operator $\mu\frac{\text{d}}{\text{d}\mu}$ to both sides of each of the
relations~\eqref{renrels}, after a small algebra we arrive at the
renormalization group equations
\beq
&&
\mu\frac{\text{d}}{\text{d}\mu}\frac{1}{\ka^2}
\,=\,
\frac{8\La}{(4\pi)^2}\,,
\label{RG-ka}
\\
&&
\mu\frac{\text{d} \La}{\text{d}\mu}
\,=\,
- \frac{2 \La^2\ka^2}{(4\pi)^2}\,,
\label{RG-La}
\eeq
which are equivalent to those obtained
in \cite{TV90,BarberoG.:1993cw}.

To solve Eqs.~\eqref{RG-ka} and~\eqref{RG-La}, we define the
dimensionless quantity $\ga = \ka^2\La$. Due to the uniqueness
of this dimensionless combination of $\ka^2$ and $\La$, the
equation for $\ga$ gets factorized,
\beq
&&
\mu\frac{\text{d}\ga}{\text{d}\mu}
\,=\,
-\,\frac{10\ga^2}{(4\pi)^2}.
\label{RG-ga}
\eeq
The solution of this equation has the standard form
\beq
&&
\ga(\mu)
\,=\,
\frac{\ga_0}{1
+ \frac{10}{(4\pi)^2}\,\ga_0\ln 
\frac{\mu}{\mu_0}}\,,
\label{sol-RG-ga}
\eeq
where $\ga_0 = \ga(\mu_0)$ and $\mu_0$ marks a fiducial energy
scale. We assume the initial values of the renormalization group
trajectories of the cosmological constant $\La_0 = \La(\mu_0)$ and
the gravitational constant $G_0 = G(\mu_0)$ as it is useful to
come back from $\ka^2$ to $G$ at this stage.

Now, using~\eqref{sol-RG-ga} in~\eqref{RG-ka} and~\eqref{RG-La},
we obtain the final solutions
\beq
G(\mu)
\,=\,
\frac{G_0\,\,\,}{
\big[1+ \frac{10}{(4\pi)^2}\,\ga_0\ln \frac{\mu}{\mu_0}\big]^{4/5}}
\label{sol-ka}
\eeq
and
\beq
\La(\mu)
\,=\,
\frac{\La_0\,\,\,}{
\big[1+ \frac{10}{(4\pi)^2}\,\ga_0\ln \frac{\mu}{\mu_0}\big]^{1/5}}\,,
\label{sol-La}
\eeq
which are certainly consistent with~\eqref{sol-RG-ga}.

The solutions~\eqref{sol-ka} and~\eqref{sol-La} are remarkable in
several aspects. First of all, such independent solutions for the two
effective charges are impossible in quantum gravity based on the usual
effective action neither in quantum general relativity nor the
fourth-derivative gravity, as the individual equations for $G(\mu)$ and
$\La(\mu)$ are completely ambiguous. In the latter model, only the
solution for the dimensionless quantity in~\eqref{sol-RG-ga} is
gauge-fixing and parametrization independent\footnote{In quantum Einstein
gravity based on the usual effective action, on the other hand, only by
using the on-shell version of renormalization group it is possible
to define an unambiguous equation for $\ga$ \cite{frts82}.}.
Here we have a well-defined running for the two parameters only
because of the use of the Vilkovisky--DeWitt effective action.

Let us note that the unambiguous solutions for $G(\mu)$ and
$\La(\mu)$ exist in the superrenormalizable gravity model
\cite{SRQG-betas}, but there are two relevant differences. The
advantage of the equations and solutions of  \cite{SRQG-betas} is
that those can be exact, in the sense of not depending on the order of
the loop expansion. On the other hand, the higher-derivative models
that lead to such an exact result imply the functional integration
over massive degrees of freedom, which can be ghosts or healthy
modes. This means that the corresponding equations are valid only
in the UV for the quantum gravity energy scale, {\it i.e.}, only in
the trans-Planckian region. Below the Planck scale the massive
degrees of freedom decouple and we are left with the quantum
effects of effective quantum gravity, such as the ones of quantum
general relativity (see \textit{e.g.} \cite{don}, the review
\cite{Burgess} and the recent discussion of the decoupling
in gravity in~\cite{Franchino-Vinas:2018gzr,324}).

On the contrary, the running described by~\eqref{sol-ka}
and~\eqref{sol-La} comes from the quantum effects of the purely massless
degrees of freedom. Up to some extent, the running
should be described by the same equations in both UV and IR.
According to the general discussion which we postpone for the
next section, these equations can be seen as exact, being valid 
in the same form even beyond the one-loop approximation. 

It is clear that the physical interpretation of the solutions
(\ref{sol-ka}) and~\eqref{sol-La} depends on the sign of $\ga_0$.
Since the positive sign of $G$ is fixed by the positive definiteness
of the theory, the sign of $\ga_0$ depends on the one of $\La_0$.
Due to the cosmological observations, we know that the sign of the
observed cosmological constant is positive in the present-day
Universe~\cite{SN-Ia-P,SN-Ia-R}. For a positive $\ga_0$  the
solutions~\eqref{sol-ka} and~\eqref{sol-La} indicate the asymptotic
freedom in the UV. In case of a moderate cosmological constant
(remember $\ka \propto M_P^{-1}$) the value of $\ga_0$ is
very small. This implies a very weak running, that is irrelevant
from the physical viewpoint. In particular, the running~\eqref{sol-ka}
and~\eqref{sol-La} is not essential for the cosmological constant
problem between the electroweak scale and the present day,
low-energy, cosmic scale.

On the other hand, at the electroweak energy scale, the early
Universe probably passed through the corresponding phase
transition. At that epoch, the observable value of the cosmological
constant could dramatically change because of the symmetry
restoration. Does this change $\La$ in the action~\eqref{action}?
The answer to this question is negative. Let us remember that the
observable cosmological constant is a sum of the two parts: one
is the vacuum parameter in the gravitational action~\eqref{action}
and another is the induced counterpart, the main part of it coming
from the symmetry breaking of the Higgs potential. The main
relations are  (see, \textit{e.g.},~\cite{Weinberg89}
or~\cite{CC-nova})
\beq
\rho_\La^{obs}\,=\,
\rho_\La^{ind}\,+\,\rho_\La^{vac},
\qquad
\rho_\La^{ind}
\,=\,
\frac{\La_{ind}}{8\pi G_{ind}}\,=\,-\,\la v_0^4,
\label{ind EH}
\eeq
where $\la$ is the self-coupling and $v_0$ the vacuum expectation
value of the Higgs field. As far as $\rho_\La^{ind}$ is negative and
the magnitude of $\rho_\La^{obs}$ is negligible, the sign of
$\rho_\La^{vac} = \frac{2\La}{\ka^2}$ is positive, independently of
the electroweak phase transition.

Thus, we conclude that the sign of $\ga_0$ is always
positive, at least between the present-day cosmic scale in the IR and
the GUT scale in the UV, where the considerations based on the
Minimal Standard Model formulas, such as~\eqref{ind EH}, may
become invalid. In all this interval, the value of $\ga_0$ is
numerically small, such that the running in~\eqref{sol-ka}
and~\eqref{sol-La} is not physically relevant.

One can imagine a situation in which another phase transition occurs
at the GUT scale (that means about $10^{14}$--$10^{16}\,\text{GeV}$),
such that the new vacuum $\La$ between this scale and the Planck
scale $M_P \approx 10^{19}\,\text{GeV}$ is negative. Then, the
solutions (\ref{sol-ka}) and~\eqref{sol-La} indicate the asymptotic
freedom in the IR.  Furthermore, if the cosmological constant in this
energy scale interval has the order of magnitude of $M_P$, these
solutions describe the situation of a dramatically strong running of
both constants $G$ and $\La$, which are strongly decreasing in the
IR. As we have learned in the previous Sec.~\ref{Sec2}, for the
values satisfying $|\La| \ll M^2_P$, the higher-loop contributions
cannot modify the form of the running. In any case, the construction
of the corresponding model of GUT would be an interesting subject to
work on in future. Here we just want to note that our results indicate
this possibility.

\section{Renormalization group for the fourth-derivative parameters}
\label{Sec4}

In order to complete the discussion, let us consider the renormalization
group equations for the fourth-derivative terms in the action of
gravity. To this end, we have to complement the action~\eqref{action}
with at least all those terms which are present in the
expression for the divergences \eqref{Final-natural}. According to the power
counting \eqref{omega}, at $p$-loop order it is necessary to introduce
into the action terms with up to $2p+2$ derivatives of the metric.
In this way we arrive
at the well-known action of the higher-derivative quantum gravity
\cite{highderi},
\beq
S_{\text{tot}}
&=&  \int \text{d}^4 x \sqrt{-g} \,
\Big\{
-  \frac{1}{\ka^2}\big(R+2\La\big)
-  \frac{1}{2\la}C^2
+ \frac{1}{2\rho}E
-  \frac{1}{2\xi}R^2
+ \frac{1}{2\ze}\,C_{\mu\nu\al\be}
\,C^{\al\be}_{\,\cdot\,\,\cdot\,\rho\si}
\,C^{\rho\si\mu\nu}
\nn
\\
& & + \,
\sum_{n=1}^{N}\Big[
\om_{n,C} C_{\mu\nu\al\be}\, \square^n\,C^{\mu\nu\al\be}
+ \om_{n,R} R \, \square^{n} R\Big]
+ {\mathcal O}(R_{\dots}^3)\Big\},
\mbox{\qquad}
\label{action-high}
\eeq
where $\la$, $\rho$ and $\xi$ are the
dimensionless parameters of the action and $N=p-1$.
The terms with more than four derivatives which contribute to the propagator
of the quantum metric have the forms  $ R \square^{n} R$ and
\beq
&&
C_{\mu\nu\al\be}\, \square^n\,C^{\mu\nu\al\be}
\,=\,
R_{\mu\nu\al\be} \square^n R^{\mu\nu\al\be}
- 2 R_{\mu\nu} \square^n R^{\mu\nu}
+ \frac13\,R \square^n R.
\label{C2n}
\eeq
One could also include the curvature-squared higher-derivative terms of the type
\beq
&&
\mathrm{GB}_{n}
\,=\,
R_{\mu\nu\al\be} \square^{n} R^{\mu\nu\al\be}
- 4 R_{\mu\nu} \square^{n} R^{\mu\nu}
+ R \square^{n} R
\label{GBn}
\eeq
which represent the extended version of the four-dimensional
Gauss--Bonnet topological invariant. Of course, these terms are not
topological for $n \geqslant 1$, but can be shown (see \textit{e.g.} \cite{highderi})
to be ${\mathcal O}(R_{\dots}^3)$ and therefore they do not contribute
to the propagator. For the sake of simplicity we assume that the
sum in (\ref{action-high}) is finite, as otherwise we arrive at the
non-local actions of quantum gravity (see, \textit{e.g.}, \cite{Modesto-nonloc}
for a review). In such a case, the structure of massive poles of the propagator
when loop effects are taken into account is more complicated~\cite{CountGhosts}
and is not relevant for the present discussion.
Still regarding  Eq.~(\ref{action-high}), we point out that
we separated
one of the possible Weyl-cubic terms $C_{...}^3$ from other
terms of third and higher-order in curvatures, because
in what follows we shall use it to discuss the two-loop
effective low-energy beta function for the parameter $\ze$.

In the polynomial theories (\ref{action-high}),
the propagator can have real massive poles \cite{highderi} or
complex ones \cite{Modesto-complex}, but in both cases the natural
situation is that all these massive parameters have the Planck order
of magnitude~\cite{Seesaw}. Thus, in the effective approach, below
the Planck scale we can completely ignore the quantum contributions
of these massive degrees of freedom. The quantum effects are coming
only from the massless mode, associated to the
Einstein--Hilbert action (\ref{action}).

The expression (\ref{action-high}) includes the action of the
fourth-derivative gravity \cite{Stelle77},
\beq
S_{\text{four}}
\,=\,  \int  \text{d}^4 x \sqrt{-g} \,
\left\{
-  \frac{1}{2\la}C^2
+ \frac{1}{2\rho}E
-  \frac{1}{2\xi}R^2
-  \frac{1}{\ka^2}\big(R+2\La\big)\right\},
\label{action-tot}
\eeq
as a particular case.
At the one-loop level, the power counting shows that only the terms up to the
four-derivative part of the action (\ref{action-tot}) gains divergent
contributions and, correspondingly, receives the logarithmic
non-local corrections.
Thus, we shall consider in details
the beta functions and renormalization group equations for
the remaining parameters in this sector of the total action.

The renormalization group equations for  $\la$, $\rho$, and $\xi$
were previously explored in the framework of the semiclassical theory,
starting in \cite{nelspan82} (see \cite{book,QFT-OUP} for a
formal consideration and further references), and higher-derivative
quantum gravity \cite{frts82,avbar86,Gauss}. In the effective
approach to quantum gravity based on the standard effective action
one can determine only the equation for $\rho$ since the
corresponding divergence survives on-shell, see Eq. (\ref{Final-onshell}),
being unambiguous. On the other hand, the universality of the
Vilkovisky--DeWitt effective action \cite{Vil-unicEA,DeWitt-ea} makes it possible
 the new version of the renormalization group equations for
 the parameters $\la$ and $\xi$.
To this end, one applies the standard algorithm to the
Eqs.~\eqref{Final-natural} and~\eqref{action-tot}, from which it follows the
beta functions
\beq
&&
\be_\la\,=\,-\,\frac{a_{\mbox{\tiny QG}}^2}{(4\pi)^2}\,\la^2\,,
\qquad
a_{\mbox{\tiny QG}}^2\,=\,\frac{121}{30},
\label{betalambda}
\\
&&
\be_\xi\,=\,-\, \frac{b_{\mbox{\tiny QG}}^2}{(4\pi)^2}\,\xi^2\,,
\qquad\,
b_{\mbox{\tiny QG}}^2\,=\, \frac{31}{18},
\label{betaxi}
\\
&&
\be_\rho\,=\,-\,\frac{c_{\mbox{\tiny QG}}^2}{(4\pi)^2}\,\rho^2\,,
\qquad\,
c_{\mbox{\tiny QG}}^2\,=\,\frac{151}{90}.
\label{betarho}
\eeq

We have to define the lowest possible IR scale. In flat spacetime,
the running produced by the
quantum effects of the massless fields can be considered to occur
for arbitrarily low energies. However, in the real applications
(even in the low-energy cosmology) there is a natural IR cut-off, as
it was described in Ref.~\cite{GWprT}. In order to understand the
origin of this cut-off, let us remember that
the running of the fourth-derivative terms is related to the
logarithmic form factors \cite{bavi90}. For the Weyl-squared term,
for example, the corresponding term in the effective action reads
\beq
\frac{a^2}{(4\pi)^2} 
\int \text{d}^4 x  \sqrt{-g}\,\,C_{\mu\nu\al\be}
\,\ln \Big(-\frac{\Box}{\mu^2_0}\Big) \,C^{\mu\nu\al\be},
\label{FF-Weyl}
\eeq
with $a^2$ defined in (\ref{RGlambda}) below;
while the corresponding d'Alembert operator for weak
perturbations around the cosmological (isotropic and
homogeneous) background has the form (see \cite{GWprT} for
the details)
\beq
\Box
=
\pa_t^2 - 4H\pa_t - 2{\dot H} - 10H^2
\,+\,\dots\,,
\label{Weyl-Box}
\eeq
where $H$ is the Hubble parameter.
The expression (\ref{Weyl-Box}) shows that in the far future
of the Universe, with the background becoming close to the
de~Sitter space,  there will not be physical running of $\la$,
because the theory is effectively massive. The cut-off (fictitious
mass) parameter is defined by the relation $\,H \sim \sqrt{\La/3}$.
Numerically, this means that the running ends in the IR at the
scale of the order $\,H_0 \approx 10^{-42}$~GeV. Between this
scale and the intermediate scale defined by the neutrino masses
(presumably of the order  $\,m_\nu \approx 10^{-12}$~GeV) the
running of $\la$, $\xi$ and $\rho$ is defined by the contributions
of effective quantum gravity (\ref{betalambda}), (\ref{betaxi}) and
 (\ref{betarho}) and the ones of the photon. Starting from the
neutrino scale, we have to include fermion contributions.
Thus, the renormalization group equation for $\la$  is
\beq
\mu\frac{d\la}{d\mu}\,=\,-\frac{a^2}{(4\pi)^2 } \, \la^2\,,
\qquad
a^2
\,=\,
a_{\mbox{\tiny QG}}^2\,+\, \frac{1}{5} \,+\, \frac{N_f}{10},
\label{RGlambda}
\eeq
where $N_f$ is the number of fermions. The solution of this equation
has the usual form
\beq
\la(\mu)
\,=\,
\frac{\la_0}{1 + \frac{a^2}{(4\pi)^2}\,\la_0\,\ln\frac{\mu}{\mu_0}}
\,,\qquad \la_0 = \la(\mu_0)\,.
\label{sol-lambda}
\eeq

The remaining equations for $\xi$ and $\rho$ are
\beq
&&
\mu\frac{d\xi}{d\mu}\,=\,-\, \frac{b^2}{(4\pi)^2} \,\xi^2 ,
\qquad
b^2
\,=\,
b_{\mbox{\tiny QG}}^2,
\label{RGxi}
\\
&&
\mu\frac{d\rho}{d\mu}\,=\,- \frac{c^2}{(4\pi)^2} \,\rho^2,
\qquad
c^2
\,=\,
c_{\mbox{\tiny QG}}^2\,+\, \frac{31}{90}
\,+\,\frac{11}{180} \,N_f.
\label{RGrho}
\eeq
It is worth noting that the photon and fermion contributions to $b^2$
are ruled out due to the conformal invariance of these two fields in
the massless versions. Another interesting point is that the contributions
of effective quantum gravity to the equations for $\la$ and $\rho$
have the same sign of the ones related to vector and fermion
fields. We remark that the same sign pattern also
takes place in the scalar field theory, in fourth-derivative quantum
gravity \cite{frts82} (see also \cite{avbar86,Gauss} for a verification)
and conformal quantum gravity \cite{antmot,frts82,Weyl}.
This universality of signs probably means there are some
general rules for the quantum corrections which we do not
understand yet.

The solutions of Eqs. (\ref{RGxi}) and (\ref{RGrho}) have the form
\beq
\xi(\mu)
\,=\,
\frac{\xi_0}{1 + \frac{b^2}{(4\pi)^2}\,\xi_0 \ln\frac{\mu}{\mu_0}}
\,,\qquad
\xi_0 = \xi(\mu_0)\,.
\label{sol-xi}
\eeq
and
\beq
\rho(\mu)
\,=\,
\frac{\rho_0}{1 + \frac{c^2}{(4\pi)^2}\,\rho_0
\ln\frac{\mu}{\mu_0}}
\,,\qquad
\rho_0 = \rho(\mu_0)\,.
\label{sol-rho}
\eeq

Let us make an important observation based on the discussion
in Sec.~\ref{Sec2}.
Since the theory is not renormalizable by power counting, the
dimensional arguments show that the higher-loop contributions
to the beta functions for the parameters $\la$, $\rho$ and $\xi$
are possible only for the non-zero cosmological constant. In this
case, because
the coupling (parameter of the loop expansion) in the
theory~\eqref{action} is $\ka^2 \sim M_P^{-2}$,
the higher-loop corrections to the fourth-derivative terms are
given by power series in the parameter
$\frac{\La}{M_P^2} \sim 10^{-120}$. Thus,
in the real physical situations the higher-loop corrections for the
dimensionless parameters $\la$, $\rho$ and $\xi$ are negligible.
This is true at least until the UV energy scale defined by the
electroweak phase transition. At a higher energy scale $M$, the
value of the induced cosmological constant density (see the discussion
in the previous section) is $\rho_\La \propto M^4$, such that
$\La \approx \frac{M^4}{M_P^2}$. Then the dimensionless
parameter of expansion in loops, in the framework of effective
theory, is defined by  the value of the ratio
$\frac{\La}{M_P^2} \approx \frac{M^4}{M_P^4}$. For the
electroweak phase transition, $M \approx 300$ GeV and this
parameter is about $10^{-66}$. Assuming another
phase transition at the GUT scale, we meet
$M = M_X = 10^{14}$--$10^{16}$ GeV and the
dimensionless parameter varies between  $10^{-20}$ and
$10^{-12}$. All these values are certainly sufficient to claim
the dominance of the one-loop effects. Therefore,  the running
which we have just derived, based on the effective approach to
quantum general relativity within the Vilkovisky--DeWitt formalism,
can be safely regarded as the exact, nonperturbative effect.
Indeed, the same also
concerns the running of  $\La$ and $G$, which were discussed in
the previous section.

Compared to other models of quantum gravity, the same
level of generality can be achieved only in the polynomial
\cite{highderi,SRQG-betas} and non-local models of quantum
gravity (see the power-counting discussion in \cite{CountGhosts}),
which are super-renormalizable. In both cases the beta functions
for  $\la$, $\rho$ and $\xi$ are not present in the published literature,
and for the latter it is not clear how those functions can
be derived, at least in a covariant way. Moreover,
if the
massive degrees of freedom in these super-renormalizable models
have masses of the Planck order of magnitude, the exact running
occurs only in the trans-Planckian region. On the contrary, in the
case under discussion here, the exact running is an IR effect, taking
place only {\it below} the Planck scale.

It is also worthwhile to make a comparison with the non-perturbative
analysis in quantum gravity based on the functional renormalization
group (see, \textit{e.g.}, the reviews \cite{NiederReuter,Percacci-2007} and
the more recent \cite{FRG-QG-review}). It was recently shown that
in this approach the effective average action remains gauge-fixing
dependent even on-shell\footnote{For a previous discussion
on this subject in the context of Yang--Mills fields, see~\cite{FRG-Lavrov-2012}.} \cite{BLRT}.
For this reason, even within the Vilkoviksy--DeWitt formulation of
the off-shell effective action, regardless the last being gauge-fixing
independent by construction, the effective {\it average} action
remains  gauge-fixing dependent in this case. No unambiguous
physical predictions can be extracted from the quantum calculations
in this approach.

The scheme of deriving the beta functions for $\la$, $\rho$ and $\xi$
described above resembles more the running of the vacuum action
parameters in the semiclassical gravity
\cite{nelspan82,tmf,Toms83,book}
than the renormalizable gravity \cite{frts82}. The similarity with
the semiclassical case is based on the fact that the running occurs
in the sector of the theory which does not define the quantum
effects. In the present case, this sector is related to fourth-derivative
terms in the action (\ref{action-tot}).
At higher loops one can meet the renormalization group running
for the parameters of six- and higher-derivative terms in the action
(\ref{action-high}).

As a further illustration of the method, let us derive the two-loop
beta function for the unique two-loop divergence derived until now
\cite{GorSag,vanVen}, namely for the $C^3$-term in the total
action (\ref{action-high}),
\beq
\bar{\Ga}^{(2)}_{\text{div}}
\,=\,
\frac{\mu^{n-4}}{(4\pi)^4(n-4)}
\, \frac{209}{1440} \, \ka^2
\int \text{d}^n x \sqrt{-g} \, \,C_{\mu\nu\al\be}
\,C^{\al\be}_{\,\cdot\,\,\cdot\,\rho\si}
\,C^{\rho\si\mu\nu}\,.
\label{GSvV}
\eeq
Using the standard routine, we arrive at the beta function for the
parameter $\ze$ in the action (\ref{action-high}),
\beq
&&
\be_\ze\,=\,-\,\frac{a_{W3}^2}{(4\pi)^4}\,\ka^2\,\ze^2\,,
\qquad
a_{W3}^2\,=\,\frac{209}{720}.
\label{betaze}
\eeq
We shall skip the discussion of possible matter-gravity contributions
in this case and restrict the consideration to the pure quantum
gravity model, where the results are available. In the effective
quantum theory with $\La=0$, the expression
(\ref{betaze}) is exact, while  in the case $\La \neq 0$ it gains
third- and higher-loop corrections in the form of a power series in
$\frac{\La}{M_P^2}$. As we already discussed above, in the
physically relevant situations these contributions are strongly
suppressed compared to the leading two-loop term.

It is interesting to notice that the beta function~\eqref{betaze}
depends on $G$ (via $\ka^2$). This is a general feature that occurs
with all the divergent terms whose number of derivatives is different
than four, and it is related to the fact that the coupling $\ka$ has
negative mass dimension---or, in other words, to the non-renormalizability
of the theory. Here, however, the situation is different from the
Eq.~\eqref{RG-La} defining the running  of the cosmological constant.
In fact, the Eq.~\eqref{RG-ka} for $G$ also depends on
$\La$, but it does not depend on $\ze$. Therefore, we can use the
solution for $G$ already established in~\eqref{sol-ka} to determine the
one of $\ze$. For the other massive parameters in the total
action~\eqref{action-high} we have a qualitatively similar picture.

All in all, the running of $\ze$ between the $H_0$ scale in the IR
and the Planck scale in the UV is described by the equation
(\ref{betaze}) and the solution has the form
\beq
\ze(\mu)
\,=\,
\frac{\ze_0}{1 - \frac{a_{W3}^2}{2(4\pi)^2}
\,  \frac{\ze_0}{\La_0}
\Big[ 1 - \big( 1 + \frac{ 10 }{(4 \pi)^2} \ga_0
\ln \frac{\mu}{\mu_0} \big)^{1/5} \Big] },
\qquad
\ze_0 = \ze(\mu_0).
\label{sol-gama}
\eeq
As in the previous cases, we have chosen the sign of the term
in the action (\ref{action-high}) such that the running is the
asymptotic-freedom type in the UV for a positive value of the
corresponding parameter. Formula \eq{sol-gama} also shows that
the running of $\ze$ depends on $\ga_0$, thus the situation
is very similar to what happens with $G$ and $\La$. Since the
value of $\ga_0$ is very small the running is supposed to be
weak; the same qualitative behaviour ought to occur for the other
parameters associated to higher-order curvature terms in the total
action. The last observation is that the singularity in the limit
$\La_0 \to 0$ in the solution (\ref{sol-gama}) is explained 
because in this limit the running of $G$ does not occur within
the effective approach. In the special case $\La_0 = 0$,
Eq.~(\ref{betaze}) has the standard form of solution.

\section{Physical applications}
\label{Sec5}

Let us briefly discuss about the possible physical
applications of the
running of the parameters  $\La$, $G$, $\la$, $\rho$ and $\xi$.
Certainly it would be interesting to apply the solutions (\ref{sol-ka})
and  (\ref{sol-La}), and also the solutions for the dimensionless
parameters, to both cosmology and astrophysics. Their detailed
elaboration, nonetheless, is beyond the scope of the
present work.

First of all, the use of the running of $\La$ and $G$ requires fixing
a physical identification of the scale $\mu$ from the Minimal
Subtraction scheme of renormalization. In cosmology the most
well-motivated identification is with the Hubble
parameter, $\mu \sim H$ (see \textit{e.g.} \cite{CC-nova,Babic2002}). In astrophysics,
it was originally used the identification $\mu \sim r^{-1}$ for
 objects like stars, galaxies and their clusters, with $r$ being
the distance from the center of the object \cite{PerMerc,CC-Gruni}.
Further detailed  analysis led to the more intricate identification
of Ref.~\cite{RotCurves}, which was phenomenologically successful.
Nowadays, there are some publications on the systematic derivation
and covariant forms of the scale identification, see \textit{e.g.}
\cite{StefDom,Davi}, which enable one to apply the solution (\ref{sol-ka}).

In the case of the four-derivative terms we meet the explicit
non-local form factors given by Eq.~(\ref{FF-Weyl}) and
\beq
\frac{b^2}{(4\pi)^2} 
\int \text{d}^4 x \sqrt{-g} \, R
\,\ln \Big(-\frac{\Box}{\mu^2_0}\Big) \,R\,.
\label{FF-R2}
\eeq
An observation concerning the cosmological applications of these
two logarithmic form factors is in order. The Weyl-squared term in
the action (\ref{action-tot}) affects the gravitational wave type
cosmic perturbations, but not the background solution or density
perturbations. There are no reasons why the  numerical coefficient
of this term should have a particularly large value. Thus, the presence
of the logarithmic form factor  (\ref{FF-Weyl}) can give an effect
of the IR running, similar to what has been previously described as 
a consequence of a photon effect in \cite{GWprT}. It is remarkable
that using the unique effective action one can report on the same IR  
running in effective quantum gravity. At low energies the effect 
related to the fourth-derivative term is weak and no essential
observational manifestations should be expected. At the same time,
close to the Planckian scale, when the initial seeds of the tensor
modes of cosmic perturbations are formed, there might be some
effects of the logarithmic form factor in (\ref{FF-Weyl}). This 
issue may deserve a detailed study, but it is beyond the scope 
of the present work. 

On the other hand, the coefficient of the classical $R^2$-term in the
action (\ref{action-tot}) can be either unconstrained or fixed by
the observational data. The last is the situation in the Starobinsky
inflation \cite{star}, where one can show that this value should
be as large as $5 \,\times\, 10^8$ \cite{star83}.
In this case, even at the Planck scale, the effect of the form factor
(\ref{FF-R2}) is enhanced by eight orders of magnitude. This
situation is in sharp contrast with other models, including the
Higgs inflation and inflaton-based models, which are otherwise
equivalent to the $R^2$-based model of Starobinsky. Thus, using
quantum gravity we might gain a possibility to distinguish this
among the other inflationary models.

It is important to stress that all these expectations become possible
only because of the use of the Vilkovisky--DeWitt unique effective action.
In the usual formulation of effective quantum gravity both
beta functions associated to the terms $C^2$ and $R^2$ are
dependent on the choice of gauge-fixing and parametrization of
quantum metric \cite{JDG-QG},  preventing their use in
reasonable applications.

\section{Conclusions}
\label{Sec6}

Using the effective approach and Vilkovisky--DeWitt unique effective action
in quantum general relativity, we constructed the renormalization
group equations for the Newton and cosmological constants and for the
parameters of the fourth-derivative terms in the extended action of
gravity. The part of Newton and cosmological constants has been
considered earlier in \cite{TV90}, but our analysis is done from a
different perspective. In particular, we show that in the effective
approach all the mentioned one-loop beta functions can be regarded as
exact, meaning they do not gain significant higher-loop corrections.
The same concerns the
renormalization group equation for the coefficient of the
six-derivative term. This equation is derived on the basis of the
two-loop divergences calculated in the well-known works
\cite{GorSag,vanVen} and does not require the Vilkovisky--DeWitt approach
to be universal.

The one-loop equations come from the quantum effects of the purely
massless modes and, therefore, are valid in both UV and IR. In
the UV, the renormalization group trajectories can be used only
until the scale where the massive degrees of freedom associated to
higher-derivative terms become active. However, in the IR there are no
restrictions except the extremely low-energy Hubble scale IR cut-off.

In this respect, the renormalization group equations under discussion
strongly differ from the ones in renormalizable and
superrenormalizable models of quantum gravity. In fact, those are valid
only in the UV regime, usually with respect to the Planck scale.

\section*{Acknowledgements}

\noindent
The authors are grateful to the referee of the original version of the
work [arXiv:2006.04217] for the advice to explore the nonperturbative
aspects of the renormalization group in the Vilkovisky--DeWitt effective
action.
The work of I.Sh. is partially supported by Conselho Nacional de
Desenvolvimento Cient\'{i}fico e Tecnol\'{o}gico - CNPq under the
grant 303635/2018-5.


\end{document}